\documentclass[aps,prstper,twocolumn,groupedaddress]{revtex4-1}

\usepackage{graphicx}
\usepackage{multirow}
\usepackage{amstext}
\usepackage{amssymb}
\usepackage{bbding}

\begin{document}


\title{Secondary implementation of interactive engagement teaching techniques: Choices and challenges in a Gulf Arab context}



\author{G. W. Hitt}
\email[corresponding author: ]{george.hitt@kustar.ac.ae}
\affiliation{Department of Applied Mathematics and Sciences, Khalifa University of Science, Technology and Research, Abu Dhabi Campus, P. O. Box 127788, Abu Dhabi, United Arab Emirates}

\author{A. F. Isakovic}
\affiliation{Department of Applied Mathematics and Sciences, Khalifa University of Science, Technology and Research, Abu Dhabi Campus, P. O. Box 127788, Abu Dhabi, United Arab Emirates}

\author{M. S. Bawa'aneh \footnote{current address: Department of Physics, The Hashemite University, Zarqa 13115, Jordan}}
\altaffiliation[current address: ]{Department of Physics, The Hashemite University, Zarqa 13115, Jordan}
\affiliation{Department of Applied Mathematics and Sciences, Khalifa University of Science, Technology and Research, Sharjah Campus, P. O. Box 573, Sharjah, United Arab Emirates}

\author{N. El-Kork}
\affiliation{Department of Applied Mathematics and Sciences, Khalifa University of Science, Technology and Research, Sharjah Campus, P. O. Box 573, Sharjah, United Arab Emirates}

\author{O. Fawwaz}
\altaffiliation[current address: ]{Science Department, The GLENELG School of Abu Dhabi, P.O. Box 11877, Ruwais, UAE}
\affiliation{Department of Applied Mathematics and Sciences, Khalifa University of Science, Technology and Research, Abu Dhabi Campus, P. O. Box 127788, Abu Dhabi, United Arab Emirates}

\author{S. Makkiyil}
\affiliation{Department of Applied Mathematics and Sciences, Khalifa University of Science, Technology and Research, Sharjah Campus, P. O. Box 573, Sharjah, United Arab Emirates}

\author{I. A. Qattan}
\affiliation{Department of Applied Mathematics and Sciences, Khalifa University of Science, Technology and Research, Sharjah Campus, P. O. Box 573, Sharjah, United Arab Emirates}

\date{\today}

\begin{abstract}
We report on efforts to design the ``Collaborative Workshop Physics'' (CWP) instructional strategy to deliver the first interactive engagement (IE) physics course at Khalifa University of Science, Technology and Research (KU), United Arab Emirates (UAE). To these authors' knowledge, this work reports the first calculus-based, introductory mechanics course on the Arabian Peninsula using Physics Education Research (PER)-based instruction. A brief history and present context of general university and STEM teaching in the UAE is given. We present this secondary implementation (SI) as a case study of a novel context and use it to determine if PER-based instruction can be successfully implemented far from the cultural context of the primary developer and, if so, how might such SIs differ from SIs within the US, in terms of criteria for and risks to their success. With these questions in view, a pre-reform baseline comprised of Maryland Physics Expectations in Physics (MPEX) survey, Force Concept Inventory (FCI), course exam grades and English language proficiency data are used to design a hybrid implementation of Cooperative Group Problem Solving (CGPS).  We find that for students with high English proficiency, normalized gain on FCI improves substantially, from $\left\langle g\right\rangle =0.16\pm0.10$ pre-reform to $\left\langle g\right\rangle =0.47\pm0.08$ in the CWP pilot (standard errors), indicating a successful SI. However, we also find evidence that normalized gains on FCI are strongly modulated by language proficiency and discuss likely causes. Regardless of language ability, problem-solving skill is also substantially improved and course DFW rates are cut from 50\% to 24\%. In particular, we find evidence in post-reform student interviews that prior classroom experiences, and not broader cultural expectations about education, are the more significant cause of expectations that are at odds with the classroom norms of well-functioning PER-based instruction. We present this result as evidence that PER-based innovations can be implemented across great changes in cultural context, provided that the method is thoughtfully adapted in anticipation of context and culture-specific student expectations. This case study should be valuable for future reforms at KU, the broader Gulf region and other institutions facing similar challenges involving SI of PER-based instruction outside the US.
\end{abstract}

\pacs{}

\maketitle


\section{introduction\label{sec:introduction}} 

The use of interactive-engagement (IE) instructional strategies and curriculum resources developed through physics education research (PER) \cite{mcder1999} in North America and Europe has produced improved student problem-solving performance and deeper conceptual understanding relative to lecture-centered instruction (e.g in introductory mechanics \cite{hake1998}). More recently, increased attention has been given to the complications, and their mitigation, arising during secondary implementations (SI) of PER-based curricula\cite{belc2003} and to institutionalizing successful PER-based reforms. Specifically, evidence presented in recent studies\cite{fink2005a,fink2005b,hend2007,danc2010} show that the broader contexts in which an IE course is implemented is, for the success and sustainability of the implementation, at least as important as how well PER-based learning tasks are executed in the classroom. These broader contexts can include the departmental, institutional, student and faculty idio-/ethno-cultural contexts \cite{fink2005a}. Several broad research questions are raised, given these demonstrations of the importance of context. Specifically, how far away from the context of the developing institution can a PER-based instructional strategy be implemented? If one of the broader contexts mentioned above is very different to that of the original developing institution, are there criteria on these contexts that can help faculty who are planning a SI to predict possible risks for their reform project? Following from this, in terms of the implementation, how and to what extent can the original instructional strategy be changed in anticipation of these failure risks, to better match the contexts of the implementing institution, without compromising core functions and principles of that strategy?

The present work contributes answers to these questions by reporting results from a studio-format, hybrid implementation of cooperative group problem-solving (CGPS) \cite{hell1992a,hell1992b,CGPS1999}, University of Washington-styled tutorials \cite{UWTutorials}, and our own experimental design mini-labs in a United Arab Emirates (UAE) context at Khalifa University of Science, Technology and Research (KU hereafter). Motivated by Refs. \cite{danc2010,goldberg2010}, this work also presents a design-based methodology for choosing and changing these instructional strategies based on an analysis of the cultural expectations of its users (students), and presents a post-analysis of the strategy's efficacy.

This work is structured as follows. In Sec. \ref{sec:uaecontext}, a brief overview of the UAE and KU contexts is given, emphasizing historical and present-day status of higher education in UAE society. In Sec. \ref{sec:methodology}, the case-study methodology taken by this work is outlined. In Sec. \ref{sec:theory}, we describe a simple theoretical perspective for this work, following that of Gaffney, Gaffney, \& Beichner\cite{gaff2010}, stated concisely that educational context includes student expectations about the nature of instruction, interactions and learning and that those expectations, accommodated (or violated), contribute causally in the performance of a SI through the readiness with which students adopt or reject PER-based classroom norms. In Sec. \ref{sec:dataanalysis}, data gathered to support this study's findings are presented. These data include measures of student values and expectations, of learning gains in physics, prior-to and following our course reform project instruction, International English Language Testing Service (IELTS) test \cite{IELTS} data of English proficiency, course exam data, and student interview data gathered post-reform. In Sec. \ref{sec:designcwp}, the baseline assessment from Sec. \ref{sec:dataanalysis} and the broader contextual factors from Sec. \ref{sec:uaecontext} and \ref{sec:theory} are synthesized to create criteria for the reform project. These criteria were used to evaluate eight well-known and well-documented PER-based innovations, resulting in a selection of CGPS for implementation, and to guide modifications for adapting that instructional strategy for the KU context. In Sec. \ref{sec:discussion}, we return to the three main research questions, as listed above, and discuss their answers in light of these results. We offer concluding remarks in Sec. \ref{sec:conclusion} on the efficacy of the reform, new questions raised by this work, and consequent directions for future research.

\section{UAE Context \label{sec:uaecontext}} 

\begin{table*}
\caption{Salient demographic and learning gains data, comparing U.S. and KU engineering student populations taking traditional, calculus-based introductory mechanics.}
\begin{tabular}{ccc}
U.S. Engineering Students\cite{Atma2010} & KU Direct Admission & KU Conditional Admission\tabularnewline
\hline 
\hline 
--- & 15\% KU population & 85\% KU population \tabularnewline
19.5\% women & 70\% women & 35\% women\tabularnewline
82\% Caucasian or Asian & 80\% other MENA%
\footnote{Middle East and North Africa%
} & 90\% Gulf Arab (UAE)\tabularnewline
13\% under-represented minorities & 20\% African or Asian & 10\% African or Asian\tabularnewline
1-in-5 are 1st generation & comparable to U.S. & 1-in-3 are 1st generation\tabularnewline
in their family to attend college  &  & in their family to attend high school\tabularnewline
1-in-9 not a U.S. citizen & 9-in-10 not UAE citizens & 1-in-10 not UAE citizens\tabularnewline
17\% are ELL students & $>$90\% are ELL students & 100\% are ELL students\tabularnewline
8.0 IELTS is `native' speaker & 6.5 (89) average IELTS (TOEFL) & 5.7 (60) average IELTS (TOEFL)\tabularnewline
FCI/FCME $\left\langle g\right\rangle _{\text{{Hake}}}\sim0.22\pm0.02$\cite{hake1998calcbased} & $\left\langle g\right\rangle _{\text{{Hake}}}=0.20\pm0.05$ & $\left\langle g\right\rangle _{\text{{Hake}}}=0.03\pm0.03$\tabularnewline
FCI/FCME pre-test gender gap $\sim13\% \pm 5\%$ & $14\% \pm 4\%$ & $5\% \pm 2\%$ \tabularnewline
FCI/FCME $\left\langle g\right\rangle _{\text{{Hake}}}$ gender gap $\sim0.06\pm0.05$ & $0.13\pm0.02$ & $0.02\pm0.01$ \tabularnewline
DFW rate:$10-20\%$ \cite{beichner2008} & DFW rate: 7\% & DFW rate: 50\%\tabularnewline
\hline
\end{tabular}
\label{demographics}
\end{table*}

Major political and economic changes in the Middle East and North Africa (MENA) region often initiate or come in tandem with large-scale educational reforms \cite{worldbank2008}. See Refs.\cite{heard2005,davi2008} for further review. Education in the lower Gulf coast of the Arabian Peninsula is no exception and has undergone several rapid changes in recent history. Prior to the 16th century, the lower Gulf coast's economy was mostly subsistence and did not permit the labor specialization necessary for widespread, formal education of the general public. During the middle decades of the 19th century, however the first formal schools appeared, funded by a boom in commercial pearling revenues. When that industry collapsed during the Great Depression, large-scale, domestic formal schooling mostly disappeared as well. The ensuing hardships and lack of broad access to education persisted even after the discovery of oil in the Trucial States territory in October, 1969. 

On December 2, 1971, the seven hereditary monarchies in the Trucial States region of Abu Dhabi, Ajman, Dubai, Fujairah, Ras al-Khaimah, Sharjah, and Umm al-Qaiwain declared their formation of the United Arab Emirates. Concurrently, the UAE Ministry of Education, along with many other federal ministries, were created to oversee a new public school system, using curricula imported mainly from Kuwait and Jordan and primarily teacher-driven, rote-learning methods, as textbooks and other resources were not yet widely available. The growth of oil revenues, beginning in the 1970's, however lead to huge expansions in affluence and access to education for the region. Over the span of a few generations, UAE society rapidly transformed from one of about 80,000 Gulf Arabs, with a per capita income of 3K USD (2005 dollars) and an adult literacy rate of $<$ 10\%, to one that at present has nearly 6 million people, with expatriate groups from 90 nations, a per capita income of 33K USD, and an adult literacy rate among citizens of $~$80\%.

At present, there are 19 institutions of tertiary education in the Emirate of Abu Dhabi alone, including KU. A few salient features of the current higher education landscape are as follows \cite{nridge}. Combined, these institutions have a gross enrollment of about 25-30\% of the adult citizen population, a factor of 5 increase over that of 1970 and 75\% of which are female students. The language of instruction in most settings is English. Consequently, while each institution has a distinct core mission, all that teach in English share a need to accommodate a majority of English Language Learners (ELLs), graduating from the mostly Arabic-based secondary schools. For these students, most institutions have a language-conditional admission category and a ``foundation'' or ``preparatory'' program (see Demaree \emph{et al.}, 2008 as an example\cite{demar2008}), a year-long, intensive English and remedial math-and-science curriculum. As a result, the average time spent studying to obtain a bachelor's degree is 5.5 years. Another ongoing challenge, especially for STEM-focused programs, is the relatively small number of students following science and mathematics-intensive tracks in secondary school ($<$ 5\%) and selecting to study STEM disciplines at university ($<$ 30\%). 

KU was established in 2007 by royal decree and by acquisition of Etisalat University College in Sharjah, UAE which now forms its Sharjah campus facility. The Abu Dhabi campus opened its doors to degree program students in Fall 2009. Currently, KU is composed of the College of Engineering only which includes the Department of Applied Mathematics and Sciences where its mathematics and natural sciences faculty are employed. All students, about 1000 total, are engineering majors who take two calculus-based introductory physics courses delivered by the department. Across the two campuses, these two courses currently serve 250-300 students per year, about 85\% of whom are UAE nationals. 

Table \ref{demographics} summarizes salient demographics comparing US and KU engineering student populations and the performance of traditional instruction for introductory mechanics in terms of conceptual learning gains, conceptual gender gap, and course drop-fail-withdrawal (DFW) rate. Clearly, the KU population as a whole is substantially different from the US in these terms. Notably, the predominance of ELLs and the representation of women are remarkable. Furthermore, a detailed analysis\cite{earlyCWP} reveals equally substantial differences within the KU population, captured entirely by the admissions category. For directly admitted students, response to traditional pedagogy appears basically consistent with US students. Unfortunately, the conditionally admitted (after completing an average 2-semester ``preparatory'' instruction), UAE national majority show no statistically significant conceptual learning gain and have alarmingly high course DFW rates. Remedying this is a major underlying motivation for exploring SI of PER-based instruction at KU.

\section{methodology\label{sec:methodology}} 

A case study methodology is followed for answering the research questions of this work. The case study is constructed as follows. First, the KU student population is characterized, with the goal of understanding performance and the values and expectations about learning that the average student would bring to a reformed version of the introductory mechanics course. Existing data in the literature and new measurements of expectations, at multiple levels of context, are used to determine the degree to which broader UAE culture values manifest in UAE student expectations in specifically educational and physics instruction contexts. Specific needs for developing conceptual and problem-solving competency are identified.

Conceptual understanding and development is measured with the Force Concept Inventory (FCI) \cite{hest1992a} pre- and post-instruction over several semesters of traditional instruction pre-reform, to establish a baseline, and post-reform, to measure efficacy. Problem-solving ability is measured with course exam data, also gathered pre-reform for benchmarking, and post-reform, to measure efficacy. We triangulate on our conclusions with qualitative data gathered from post-instruction student interviews.

\section{Theoretical Perspective \label{sec:theory}}  

Our theoretical perspective in this study is formed by two key concepts. First, we understand the failure mechanisms for cross-cultural SI to include strong contributions from \emph{expectancy violation} in the creation of reformed classroom norms, as described by Gaffney \emph{et al.}\cite{gaff2010} and references therein. The classroom norms critical to well-functioning PER-based instruction are \emph{student-student interaction, hands-on activities, equality in teacher-student interactions}, and \emph{sense-making over answer-making}\cite{danc2010,turpen2010}. Second, we distinguish between differing kinds of and differing origins for expectations in terms of the \emph{frame}\cite{fink2005a,fink2005b} or \emph{level of context} where they manifest.  

To illustrate the interaction of these two concepts with an example, consider two students ``A'' and ``B'' who expect instructors to provide complete procedures for learning activities, and the frame of two example contexts, at the ``situational-'' and ``task-level'' frames; a mechanics laboratory experiment and length measurement, respectively.  At the situation level, both students expect that the instructor should provide a procedure for `doing' the experiment. At the task-level, making a length measurement, a similar expectation may or may not manifest. Despite their general expectation, neither student likely expects to be shown how to use a meter stick. Thus, context can change how an expectation is manifest. Regarding violations, if an instructor asks the student to measure the length of a table, but provides no procedure, there would be no violated expectations. However, if the instructor asks the student to design their own experiment to answer an open-ended question about projectile motion, both students' expectations will be violated. Depending on a potentially large number of other contextual factors (e.g. instructor's gender or age, having to perform tasks alone or in a team, etc.) the student will have an affective response, judging the violation negatively or positively \cite{gaff2010}. As Gaffney \emph{et al.} review and demonstrate, negative violations at the situational-level have been shown to lead to lower student engagement and decreased learning\cite{gaff2010}, similar to the manner in which negative violations of task-level expectations (e.g. on the nature of science, physics, learning, etc.) have been shown to be causal in learning gains\cite{cole2005,adam2006,korte2007}.

Furthermore, the importance for us of contextual frames as a basis, extends beyond categorizing manifestations of values and beliefs as expectations in that context. We also refer to this model to distinguish origins of such expectations. Suppose an important difference between student-A and student-B is that A has taken many laboratory courses in the past and B has a broad, culturally grounded belief that \emph{teachers are authority figures who possess and distribute absolute truths}. The origin for their expectations differ in terms of contextual frame, in that student-A's expectation is the accumulated result of repeated prior experiences at the situational-level, whereas student-B's expectation is the manifestation of a cultural value that the educational situation inherits from broader frames in which it is nested. These students would also then differ in terms of how changes in context change their expectations. If both graduate and go to work for an engineering firm, student-A might no longer expect procedures to be provided by a supervisor if the work situations differ significantly from a classroom laboratory. But the same expectation may be persistent in the case of student-B, since both educational and work situations, are nested in the same broader cultural context.

\subsection{A Novel Failure Mechanism for SIs}

In light of this perspective, we consider a potential failure mechanism for SIs of PER-based instruction that is different, by way of extension, from that suggested by Gaffney \emph{et al.}\cite{gaff2010}. National culture values, and the expectations they engender about education and learning in the idio-cultural/institutional and situational frames of context may create conditions where the adoption of critical PER-based classroom norms is not possible. Gaffney \emph{et al.} suggest a potential underlying cause for cases of failed SIs in the PER literature (all within the US, e.g. as discussed by Hake\cite{hake1998,hake1998IEM2b}) is a failure to minimize and/or manage negative expectancy violation, caused mainly by ``classroom and instructor factors that may or may not deviate from what students expect based on previous classes they have taken''\cite{gaff2010}. Our perspective extends this argument about cause of SI failure, to include scenarios where student expectations, created not by prior classroom experiences within the same culture, but those fostered by differing cultural values that conflict with and inhibit the formation of necessary PER-based classroom norms. Both mechanisms should be able to cause negative expectancy violation, rejection of PER-based norms, and SI failure. Given the UAE context described above, both failure mechanisms are likely important, confounding influences in the KU student population. To isolate the influence of the cultural mechanism, it is then important to gauge the degree to which cultural values manifest as expectations and effect engagement in the nested situational and task-level frames for the reformed course. Otherwise, there is no way to distinguish the two causes in the event of a failure: (1) failure to establish PER-norms due to negative violation of expectations formed on the basis of prior experience (likely working in isolation in a US SI), and (2) failure to establish PER-norms due to negative violation of expectations formed on the basis of broader cultural values.

\subsection{Review of supportive evidence in PER literature}

The primary research question of this work is to determine how far from the context of the developing institution can a PER-based instructional strategy be successfully implemented. We consider normalized conceptual learning gains as the litmus test of a successful implementation. Therefore, to scrutinize our theoretical perspective, we compile in Table \ref{tab:summarysecondaries} a selection of reported SIs for university physics course reforms for which FCI or FCME data was gathered, with particular interest in cases where PER developed in the US was implemented in another country or culture. As shown there, for the small number of cases we found in the literature, there does appear to be a difference between the improvement in post-reform normalized learning gains, when PER developed in a US institution is implemented outside the US. The average improvement to normalized learning gain $\Delta$-avg.gain is approximately 70\% of that attained (0.21) in SIs within the US. This difference in the literature raises the question: do cultural differences manifest in expectations that PER-norms violate in the classroom contexts (situational- and task-level) of a SI?

\begin{table*}
\caption{Implementations of PER instruction for university physics including FCI(FCME)\footnote{FCME scores are scaled to FCI for comparison\cite{thorn2009}} data both pre-/post-instruction \emph{and} gain pre-/post-reform. The national culture of the developing (DNC) and of the implementing institution (INC) is given.\label{tab:summarysecondaries}}
\begin{tabular}{l|c|c|c|c|c|c}
Implementation & Type & DNC & INC & avg. gain pre-reform & avg. gain post-reform & $\Delta$-avg. gain\tabularnewline
\hline
\hline
Modeling Instruction\cite{modeling} at ASU\cite{hake1998} & Primary & USA & USA & 0.26 & 0.50 & +0.24\tabularnewline
Peer Instruction\cite{mazur1997} at Harvard\cite{hake1998} & Primary & USA & USA & 0.29 & 0.56 & +0.29\tabularnewline
SDI Labs\cite{SDILabs} at IU\cite{hake1998} & Primary & USA & USA & 0.23 & 0.60 & +0.37 \tabularnewline
\hline
\multicolumn{2}{c|}{Average Primary reported by Hake\cite{hake1998}} & USA & USA & 0.26 & 0.55 & +0.30$\pm$0.04\tabularnewline
\hline
CGPS/Modeling at CalPoly\cite{hake1998} & Secondary & USA & USA & 0.25 & 0.56 & +0.31 \tabularnewline
CGPS/MBL at UM\cite{hake1998} & Secondary & USA & USA & 0.21 & 0.33 & +0.12 \tabularnewline
CGPS/MBL at OSU\cite{hake1998IEM2b} & Secondary & USA & USA & 0.13 & 0.42 & +0.29 \tabularnewline
CGPS/MBL at UL\cite{hake1998IEM2b} & Secondary & USA & USA & 0.18 & 0.26 & +0.08 \tabularnewline
CGPS/PI at UML\cite{hake1998IEM2b} & Secondary & USA & USA & 0.19 & 0.22 & +0.03 \tabularnewline
\hline
\multicolumn{2}{c|}{Average Secondary reported by Hake\cite{hake1998}} & USA & USA & 0.19 & 0.36 & +0.17$\pm$0.06 \tabularnewline
\hline
Studio Physics at CSM\cite{furt2001} & Secondary & USA & USA & 0.15 & 0.65 & +0.50 \tabularnewline
Interactive Lecture Demos. at RPI\cite{cumm1999} & Secondary & USA & USA & 0.18 & 0.35 & +0.17 \tabularnewline
Cooperative Groups at RPI\cite{cumm1999} & Secondary & USA & USA & 0.18 & 0.36 & +0.18 \tabularnewline
Open Source Tutorials at FIU\cite{goer2011} & Secondary & USA & USA & 0.24 & 0.42 & +0.18 \tabularnewline
Modeling Instruction\cite{modeling} at FIU\cite{brewe2010} & Secondary & USA & USA & 0.22 & 0.51 & +0.29 \tabularnewline
\hline
\multicolumn{2}{c|}{Average Secondary, all of the above} & USA & USA & 0.18 & 0.39 & +0.21$\pm$0.07 \tabularnewline
\hline
Inter. Lecture Demos. at USydney\cite{sharma2010} & Secondary & USA & AUL & (0.06) & (0.23) & +(0.17) \tabularnewline
Open Source Tutorials at FIU\cite{goer2011} & Secondary & USA & PUE & 0.24 & 0.36 & +0.11 \tabularnewline
Modeling Instruction\cite{modeling} at FIU\cite{goer2011} & Secondary & USA & PUE & 0.22 & 0.43 & +0.20 \tabularnewline
\hline
This work, hybrid CGPS at KU & Secondary & USA & UAE & 0.03 & 0.14 & +0.11 \tabularnewline

\end{tabular}
\end{table*}


\subsection{Cultural effects on situational expectations}
For situational expectations, Hofstede's\cite{hofs2001} seminal work on `cultural dimensions', a theoretical framework for comparative studies of national and institutional cultural values, is directly applicable for the purpose of gauging the effect of national cultural values on the manifestation of situational expectations. We focus on the first three of Hofstede's cultural dimensions constructs because they each have a direct bearing on student expectations in educational contexts. These are \emph{power distance}, \emph{uncertainty avoidance}, and \emph{individualism}. Power distance refers to ``the extent to which the less powerful members of organizations, institutions, and families accept and expect authority to be distributed unequally.''\cite{hofs2001pdi} Uncertainty avoidance refers to the ``society's tolerance for ambiguity.''\cite{hofs2001uai}  Individualism refers to ``the degree to which individuals are not integrated into groups.''\cite{hofs2001idv} Each of the dimensions are measured on a 100-point scale based on responses to the Values Survey Module\cite{VSM94}, in this case the 1994 version (VSM94). Table \ref{tab:nationalculture} below summarizes the expectations in educational situations that are reliably correlated\cite{hofs2001} with high and low scores and gives the specific scores for the nations involved in SIs of PER methods presented in Tab. \ref{tab:summarysecondaries}.

\begin{table*}
\caption{A selection of Hofstede's dimensions of national cultural values\cite{hofs2001}, VSM94 scores for a selection of nations, and expectations for educational situational norms that are reliably correlated with VSM94 scores.\label{tab:nationalculture}}
\begin{tabular}{c|c|c|c|c|c|c}
Cultural Dimension & Expectations & Expectations & \multicolumn{4}{c}{National Scores}\tabularnewline
and definition & for Low Scores & for High Scores & USA & AUL & PUE & UAE \tabularnewline
\hline
\hline

\begin{minipage}[t]{0.5\columnwidth} Power distance - the extent to which the less powerful members of organizations, institutions, and families accept and expect authority to be distributed unequally \end{minipage} & \begin{minipage}[t]{0.5\columnwidth} student-teacher equality, student-centered education, students initiating communication\end{minipage} & \begin{minipage}[t]{0.5\columnwidth} student deference and dependence on teachers, teacher-centered education, teacher-initiated communication\end{minipage} & 40 & 36 & 64 & 90 \tabularnewline

\hline

\begin{minipage}[t]{0.5\columnwidth} Uncertainty Avoidance - the society's tolerance for ambiguity\end{minipage} & \begin{minipage}[t]{0.5\columnwidth} open-ended learning situations, good discussions, tasks with uncertain outcomes that involve risk estimation and problem solving \end{minipage} & \begin{minipage}[t]{0.5\columnwidth} highly structured learning situations, teachers possessing and transferring absolute truths, tasks with sure outcomes that involve following instructions and no risk \end{minipage} & 46 & 51 & 76 & 80 \tabularnewline

\hline

\begin{minipage}[t]{0.5\columnwidth} Individualism - the degree to which individuals are not integrated into groups \end{minipage} & \begin{minipage}[t]{0.5\columnwidth} grouping according to prior affiliations (ethnic, family, friendship, etc.), no speaking out in class or in large groups, discouragement for individual initiative, and an emphasis on learning how 'to do' \end{minipage} & \begin{minipage}[t]{0.5\columnwidth} grouping according to tasks, speaking out in class or in large groups, encouragement for showing individual initiative, and an emphasis on learning how 'to learn' \end{minipage} & 91 & 90 & 88 & 25 \tabularnewline

\hline

\multicolumn{3}{r|}{Rank out-of-90 for cultural distance $(d)$, relative to USA:} & 0 & 1 & 21 & 71 \tabularnewline

\hline

\end{tabular}
\end{table*}

Hofstede points out clearly that VSM absolute scores are meaningless, but relative differences correlate reliably with differences in expectations and carry predictive power\cite{hofs2001}. Therefore, for comparison of cross-cultural SIs, we construct a simple, semi-quantitative measure of relative cultural difference from US national cultural, in terms of the power distance (PDI), uncertainty avoidance (UAI), and individualism (IDI) indices. For a national culture \emph{i}, we first calculate the relative geometric distance on these three cultural dimensions

\begin{equation}
d_{i}=\sqrt{(\Delta \text{PDI}_{\text{USA},i})^{2}+(\Delta \text{UAI}_{\text{USA},i})^{2}+(\Delta \text{IDI}_{\text{USA},i})^{2}},
\end{equation}

where USA, \emph{i} denotes the difference in the USA versus the other culture's score. We then rank the distances \emph{d} from least to greatest, to provide a measure of cultural distance from US national culture. For example, following this procedure using scores from Table \ref{tab:nationalculture} for Australian national culture (AUL) gives a rank of 1. This means that of the 90 nations scored by Hofstede using VSM94\cite{hofs2001}, none are more similar to the US than Australia, in terms of power distance, uncertainty avoidance, individualism and the associated expectations. Therefore, all other factors being equal, one predicts that SI of US-developed, PER-based strategies in US or in Australian institutions should fair no differently and produce similar improvements to measures of learning.

Conversely, one anticipates that SIs in increasingly dissimilar cultural contexts, in terms of this ranking, should become respectively more difficult and produce smaller improvements to conceptual learning gains on the basis of larger differences in expectations and greater frequency of negative violations. US-designed instructional strategies make US students uncomfortable and less likely to engage course content relative to how different the instruction is from traditional lecture. With cross-cultural SIs, one explores the effect of the `moving the other goal post': examining how students respond to US-developed IE instructional strategies as the cultural expectations of the students pre-instruction are increasingly dissimilar to that of US students. For example, from Tab. \ref{tab:nationalculture}, calculating \emph{d} and determining a ranking for Puerto Rican (PUE) and Emirati (UAE) national cultures gives 21 and 71 out of 90, respectively. Thus, for well-executed IE teaching with PER-based instructional strategies, one predicts SIs to show decreasing improvements in student learning gains with increasing rank \emph{d}, if in fact pre-instruction student expectations have a causal relationship with learning gains, as has been suggested \cite{cole2005,adam2006,korte2007}. Figure \ref{fig:gainvscontext} shows improvement to normalized conceptual learning gain, pre- versus post-reform, based on these data and plotted in order of increasing rank \emph{d}. The steady decrease in post-reform improvement shown here lends confidence that the questions of the present study are valid and worth pursuing.

\subsection{Indications of cultural effects specific to expectations in physics}

Our perspective is also supported by existing (though limited) international data on direct measures of students' expectations in physics. For example, one expects that there should be strong similarity between Hofstede's PDI construct and MPEX Independence construct, given the strong similarity of their respective descriptors\cite{hofs2001,redi1998}. Specifically, PDI should be significantly negatively correlated with favorability scores on the MPEX Independence cluster pre-instruction. Thus, the MPEX Independence cluster score should be one indicator that expectations related to PDI (student deference and dependence on teachers, teacher-centered education, teacher-initiated communication) are manifest in the situational- and task-levels in contexts that are specific to physics classrooms. Examining the PER literature, we find that Sharma, Ahluwalia, \& Sharma\cite{shar2013} have examined MPEX data taken in four nations having relatively high PDI scores (Philippines, India (HS and UG), Turkey, and Thailand). Though small, in terms of the number of nations, these MPEX Independence cluster scores and their nations' respective PDI scores indeed have a strong negative correlation ($r =-0.7$, see Sec. \ref{ssec:mpex}). This indicator provides some further confidence that the broader scope and questions of the present study are valid to pursue. Our perspective moves back to examine the wider classroom-level context and the interaction of culturally-grounded student expectations with the classroom norms that must be formed for successful PER-based instruction. Our focus on classroom-level context justifies the inclusion of UAI and IDI constructs into our measure of cultural distance \emph{d}, as these measures have a direct bearing on student-student and student-teacher interactions.

\section{Data \& Analysis \label{sec:dataanalysis}} 

\begin{figure} 
\includegraphics[scale=0.45]{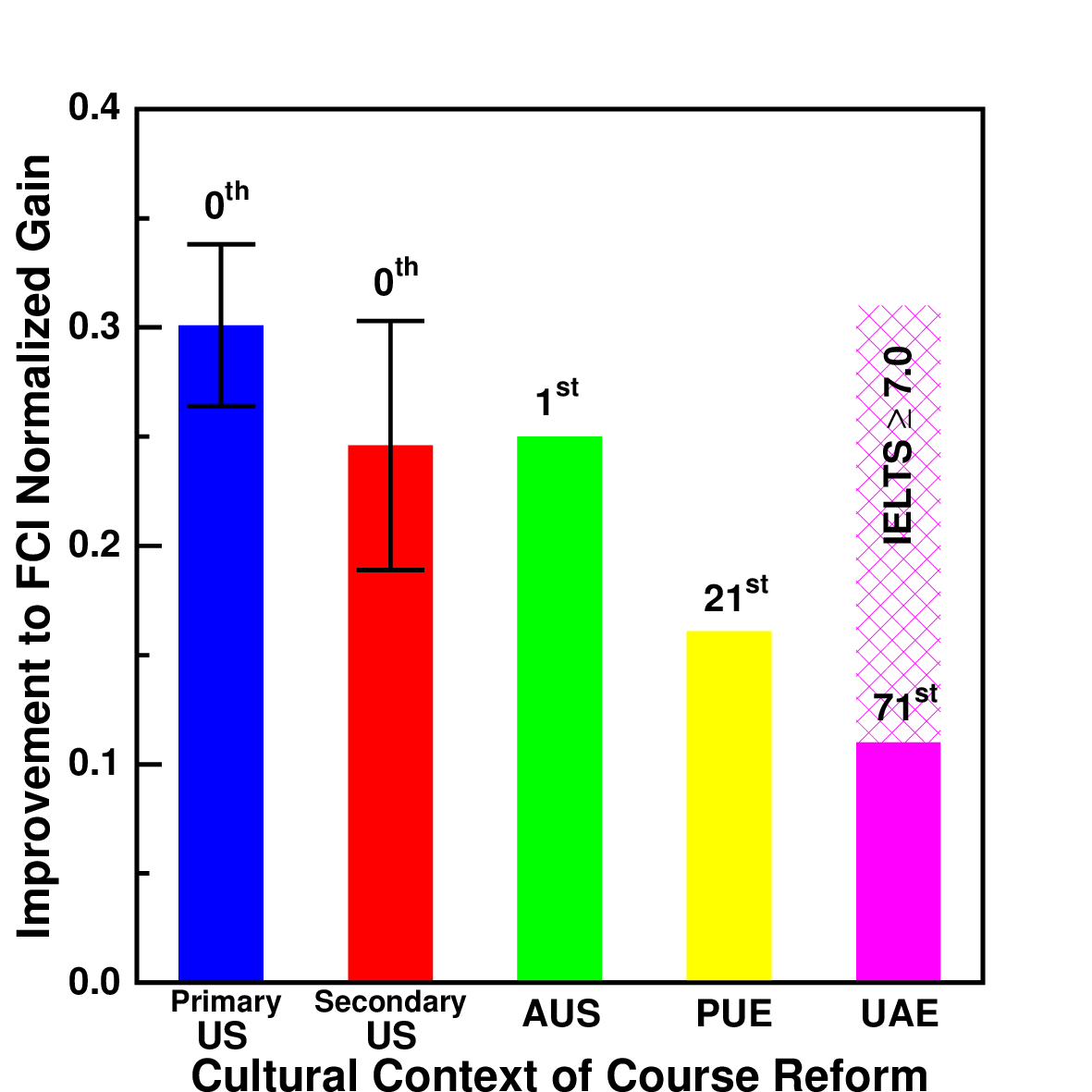}
\caption{Improvements to FCI (or FMCE) normalized gain, as a result of a PER-based course reform, for the national cultural contexts of the reforms. Error bars for Primary US and Secondary US cases are standard errors in the average, taken over the available cases in the literature. AUS, PUE and UAE represent single, SIs conducted for predominantly Australian, Puerto Rican, and Emirati student populations, respectively. Above each entry, the rank out of 79 is given for 'cultural distance' relative to US national culture, computed by calculating the geometric distance from the US score on Hostede's dimensions and ranking the results in ascending order. A low rank indicates relative cultural similarity to US national culture, in Hofstede's framework. A high rank indicates relative dissimilarity. Improvement to FCI normalized gain is replotted (purple, cross-hatched) for UAE, for only those students in the CWP reform with high English language proficiency (i.e IELTS score)  \label{fig:gainvscontext}}
\end{figure}

\subsection{Maryland Physics Expectation Survey} \label{ssec:mpex}

\begin{figure} 
\includegraphics[scale=0.5]{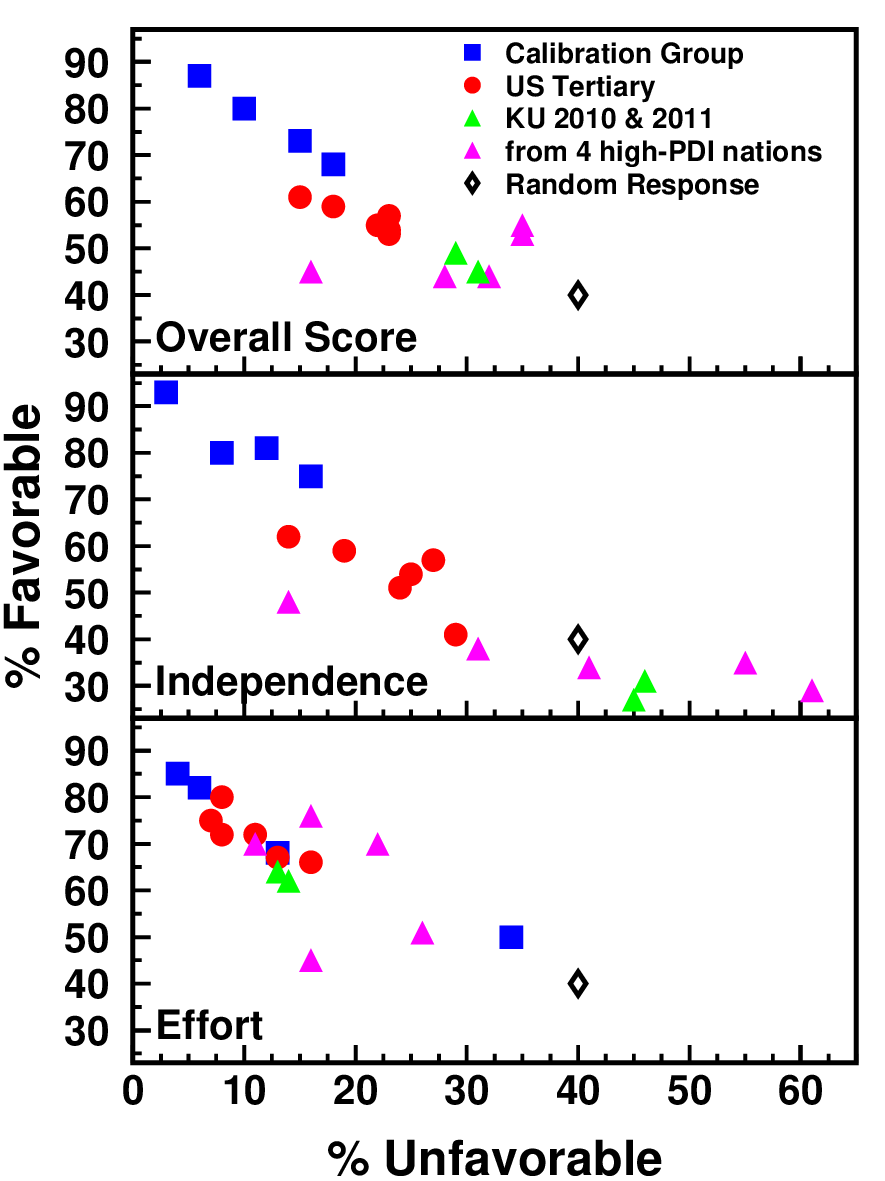}

\caption{The favorable-unfavorable Dalitz-style plot for pre-instruction MPEX survey data comparing the Calibration Group (blue squares) and US Tertiary student groups (red circles) from the original MPEX study \cite{redi1998} to KU students (green triangle), and students from data sets in other high-PDI nations (India, Turkey, Thailand, and Philippines)\cite{shar2013}. KU student data includes that for the year of (2011) and the year prior to (2010) the CWP reform. MPEX overall score (top) is compared to the question subsets for the Independence (middle) and Effort (bottom) clusters. The average for random responses (black diamond) is shown for benchmarking purposes. \label{fig:MPEX-pretest}}
\end{figure}

MPEX data was gather from KU students the year prior to the reform project (2010) and the year of the CWP reform (2011). The clusters on MPEX that arguably have the most bearing on classroom norms at the task-level are the ``Independence'' and ``Effort'' clusters (questions 1, 8, 13, 14, 17, 27 and questions 3, 6, 7, 24, 31 respectively). This is because both clusters were constructed and validated for measuring expectations about \emph{student behaviors}. As stated in Ref. \cite{redi1998}, the independence cluster measures whether or not `learning physics' ``means receiving information or involves an active process of reconstructing one's own understanding'' and for the effort cluster, ``whether {[}students{]} expect to think carefully and evaluate what they are doing based on available materials and feedback or not''. 

Figure \ref{fig:MPEX-pretest} shows the Dalitz-plot of KU student responses and those from the high-PDI set\cite{shar2013}, compared with the calibration and US tertiary groups from the original MPEX study \cite{redi1998}. Scores on the independence cluster (Fig. \ref{fig:MPEX-pretest}(middle) show the most dramatic differences with the US response and accounts for most of the overall variation. Clearly, KU students respond very unfavorably to some items in this cluster. The most unfavorable responses are to items \#1 and \#14, as was the case in the original MPEX study \cite{redi1998}, however the degree of unfavorability is significantly greater. Item \#1 states:

\begin{quote}
\emph{All I need to do to understand most of the basic ideas in this course is just read the text, work most of the problems, and/or pay close attention in class.}
\end{quote}

\noindent Only 8\% of students on average (9\% for year group 2010, 7\% for 2011) disagreed with this statement (the favorable response). No student strongly disagreed. On average 77\% of students agreed with item \#1 (76\% for KU 2010 and 78\% for KU 2011), meaning on average only 15\% of students responded neutrally. A similar pattern is present in responses to item \#14 which states:

\begin{quote}
\emph{Learning physics is a matter of acquiring knowledge that is specifically located in the laws, principles, and equations given in class and/or in the textbook.}
\end{quote}

\noindent Only 17\% of students on average (18\% for KU 2010, 14\% for KU 2011) disagreed with this statement (the favorable response). Again, no student strongly disagreed. On average 56\% of students agreed with item \#14 (59\% for KU 2010, 50\% for KU 2011), meaning on average 27\% of students responded neutrally. On the remaining items in the independence cluster, students on average responded favorably 39\% versus 35\% unfavorably and 26\% neutrally, still slightly lower but somewhat more consistent with US tertiary pre-test scores presented in Ref. \cite{redi1998}. When \emph{engaging the content} of the course, clearly, based on the text of item \#1 and \#14, KU students exhibit a particularly high dependence on authoritative sources of information (text, problems, or `in class' {[}instructor{]}) as compared to US students on average.

As shown in Fig. \ref{fig:MPEX-pretest}, KU student responses in the effort cluster are more favorable and more consistent with US patterns, though some differences remain on individual items. The notable and somewhat confusing exceptions are items \#6 and \#24 which received on average 48\% and 35\% favorable responses, respectively. Item \#6 states:

\begin{quote}
\emph{I spend a lot of time figuring out and understanding at least some of the derivations or proofs given either in class or in the text.}
\end{quote}

\noindent A 48\% agreement (the favorable response) indicates that either proofs themselves, or the contexts in which they are given (in-class or the text) is not deemed an important behavior for learning physics. Yet, item \#7 states:

\begin{quote}
\emph{I read the text in detail and work through many of the examples given there.}
\end{quote}

\noindent The level of agreement (the favorable response) is 70\% on average. Taken together, it would appear that KU students believe that engagement with the course textbook is important for learning physics, but it is most valuable as a source of worked examples, not as a source of proofs and derivations. This is somewhat in contradiction to the instructor anecdotes, that many students do not read the course text at all. It might be the case however, that the discrepancy is the result of the wording of item \#6 which contains many compound constructions (``figuring out \emph{and} understanding'', ``derivations \emph{or} proofs'', ``in class \emph{or} in the text''), that ELL students may not be sure what the statement is actually asking them to reflect upon. In support of this possibility, the Cronbach's alpha score for the effort cluster is 0.60, but if item \#6 is removed, it improves the most for any 1-item removal, to 0.70.

Item \#24 is also responded to relatively unfavorably, with only 35\% agreeing. It states:

\begin{quote}
\emph{The results of an exam don't give me any useful guidance to improve my understanding of the course material. All the learning associated with an exam is in the studying I do before it takes place.}
\end{quote}

\noindent This response is strangely at odds with student responses to item \#31:

\begin{quote}
\emph{I use the mistakes I make on homework and on exam problems as clues to what I need to do to understand the material better.}
\end{quote}

\noindent to which 83\% of students agree, the most favorable of any in the effort cluster. This discrepancy could be the result of the long tradition in UAE public schools to follow a British-style model for testing, where learning is assessed by a single, high-stakes exam at the end of the course (often carrying 60\% or more of the course's grade weight), as opposed to the model in typical US physics courses which feature 2-3 midterm exams and weekly graded homework spread throughout the course. In the prior case, with the course completed, there would be little reason for a student to expect to need to study their mistakes on an exam since the course is finished. Confirming this however, requires a more specific investigation.

\subsection{Force Concept Inventory and Course Exams}

\begin{figure} 
\includegraphics[scale=0.5]{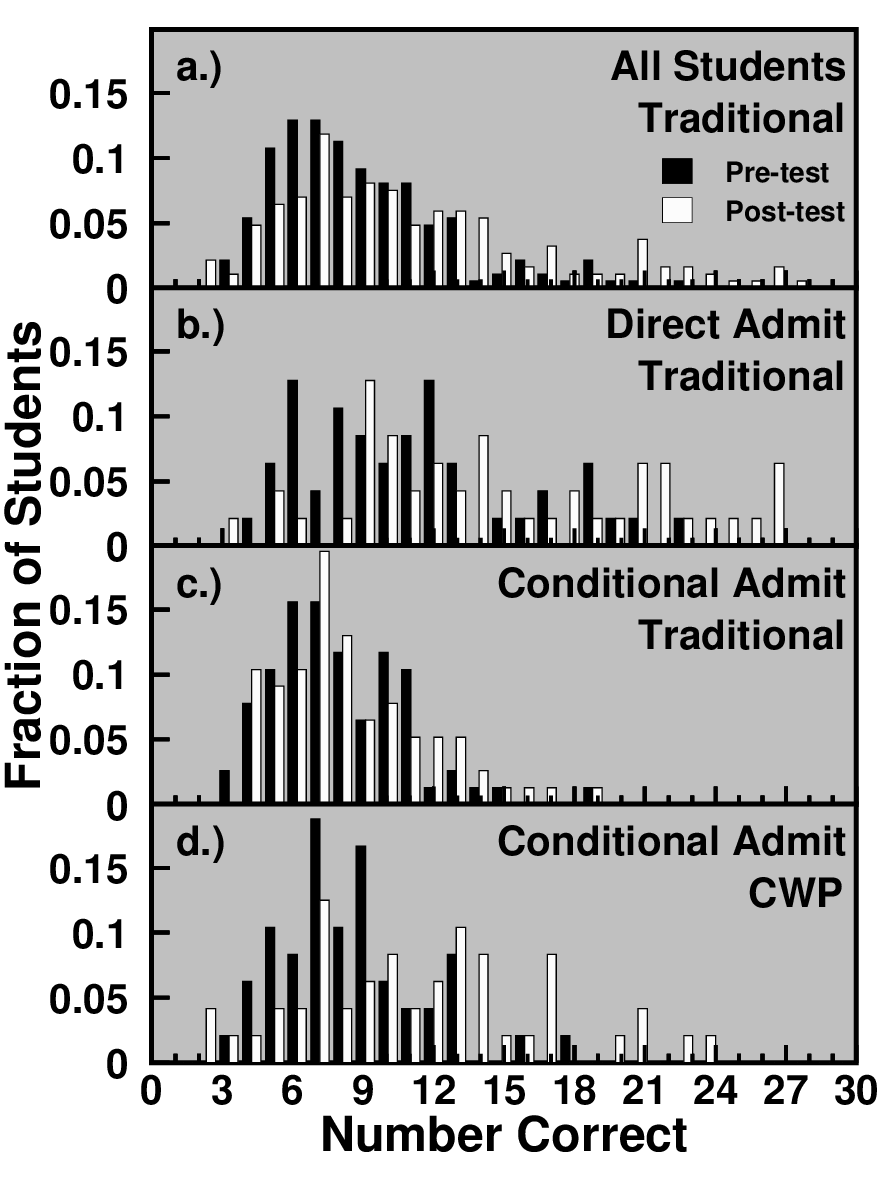}

\caption{Histogram of pre- (black) and post-test (white) Force Concept Inventory data. a.) FCI data from all traditionally taught students. b.) Data from all traditionally taught, directly admitted students. c.) Data from all traditionally taught, conditionally admitted students. d.) Data from all conditionally admitted students taught in the CWP pilot. \label{fig:fci4panel}}
\end{figure}

Starting from Fall 2009, the Force Concept Inventory (FCI) \cite{hest1992a} was administered to students both pre-/post-instruction. This was repeated in two subsequent semesters, Spring 2010 and Fall 2010, where the traditional course was offered, however logistical issues prevented administration of the Spring 2010 FCI post-test. In Fall of 2010 and Spring 2011, students had a choice of taking FCI in Arabic or English but this had no significant effect on average scores, pre- or post-instruction, over Fall 2009 or Spring 2010. Figure \ref{fig:fci4panel} shows the distribution of all FCI data considered in this work. Comparing Fig.\ref{fig:fci4panel}(c) with (d) shows clear, increased positive movement in post-test centroid. Table \ref{FCI basic analysis} shows class-averaged pre- and post-instruction FCI scores for all semesters that data have been gathered. A strict matching condition has been applied, such that any student that did not complete both pre- and post-tests is removed from the dataset prior to analysis. The uncertainties quoted for pre-test, and post-test scores are errors in the mean ($\sqrt{\sigma^{2}/N}$). The uncertainties for the Hake's gain scores are propagated from the uncertainties for mean pre-test and mean post-test scores, where the Hake's gain score \cite{hake1998} itself is calculated in the usual way as

\[
\left\langle g\right\rangle _{\textrm{Hake}}=\frac{\left\langle \textrm{Post-test}\right\rangle -\left\langle \textrm{Pre-test}\right\rangle }{100-\left\langle \textrm{Pre-test}\right\rangle },
\]
and where $\left\langle \right\rangle $ symbols indicate class-averaged
scores.

\begin{figure} 
\includegraphics[scale=0.5]{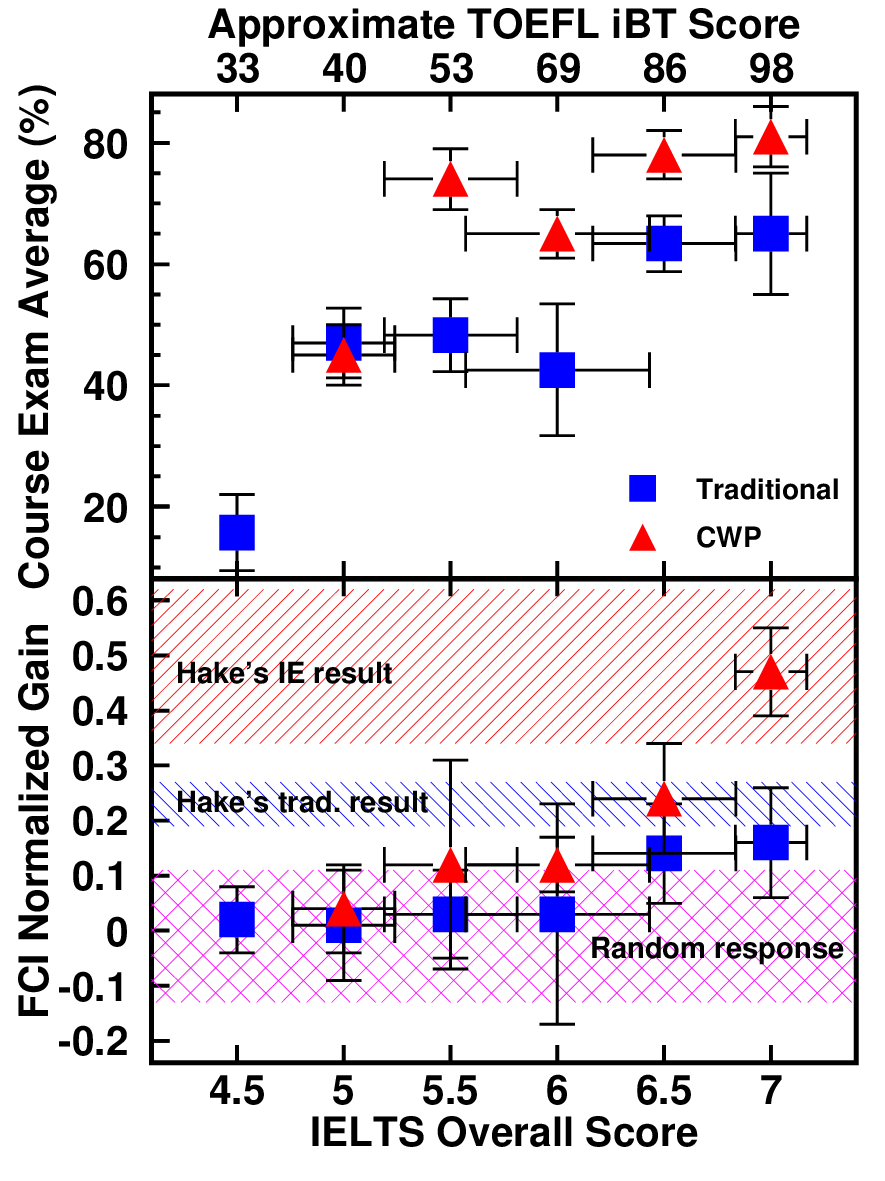}

\caption{Course exam averages (top) and FCI normalized gains (bottom) plotted versus language proficiency, as measured by IELTS overall score (bottom horizontal axis). Vertical error bars are standard errors in the mean for these data, respectively. An approximate TOEFL iBT score adapted from Table 7 of Ref.\cite{IELTS} and displayed (top horizontal axis). Horizontal error bars are approximated by a linear fit (\emph{R=0.978}) to this table\cite{IELTS} and do not necessarily represent the uncertainity in student language proficiency, but rather, the  relative uncertainty between IELTS and TOEFL iBT scales. For FCI normalized gains (bottom), Hake's result\cite{hake1998calcbased} for the average and standard error interval for traditional (blue, left-hatched) and IE (red, right-hatched) pedagogy are highlighted. The average and standard error interval for random responses (purple, cross-hatched) is highlighted for benchmarking purposes. \label{fig:examfcivslang}}
\end{figure}

Figure \ref{fig:examfcivslang} shows course exam averages (top) and FCI normalized gains (bottom) plotted versus student language proficiency, as measured by IELTS overall score (bottom horizontal axis). Vertical error bars are standard errors in the mean for these data, respectively. To aid international comparisons, the corresponding approximate TOEFL iBT score adapted from Table 7 of Ref.\cite{IELTS} and displayed (top horizontal axis). Horizontal error bars are approximated by a linear fit (\emph{R=0.978}) to this table\cite{IELTS} and do not necessarily represent the uncertainity in student language proficiency, but rather, these represent the  relative uncertainty between IELTS and TOEFL iBT scales. For course exams (top), exam data from Fall 2009 and Spring 2011 (CWP) are presented. The exams were different, so the scores are not normalized to a common scale. However, a small group of faculty not affiliated with the project were asked to conduct an item categorization (by topic) and ranking (by difficulty) task with questions from both exams. The result was that topical coverage between the two exams were similar and CWP questions were consistently judged as more difficult, so differences in Fig.\ref{fig:examfcivslang}(top) represent conservative limits on traditional problem-solving improvement. For FCI normalized gains (bottom), Hake's result\cite{hake1998calcbased} for the average gain and standard error interval for traditional (blue, left-hatched) and IE (red, right-hatched) pedagogy are highlighted. The average gain and standard error interval for random responses (purple, cross-hatched) is highlighted for benchmarking purposes. In the limit of high language proficiency, normalized gain for both traditional and CWP instruction at KU converges on the average values from Hake\cite{hake1998calcbased}. For these students, normalized gain on FCI improves substantially, from $\left\langle g\right\rangle =0.16\pm0.10$ pre-reform to $\left\langle g\right\rangle =0.47\pm0.08$ in the CWP pilot.

\begin{table*}
\caption{Force Concept Inventory data spanning Fall 2009 until Spring 2011
semesters for KU's first-semester, calculus-based physics course.}

\begin{tabular}{l|r|c|c|c|c}
Semester (Campus%
\footnote{AD is for the Abu Dhabi campus, Shj is for the Sharjah campus.%
}:Mode%
\footnote{T is for Traditional and CWP is for Collaborative Workshop Physics.%
}) & Population & Size ($N$) & FCI $\left\langle \text{{Pre-test}}\right\rangle $ & FCI $\left\langle \text{{Post-test}}\right\rangle $ & $\left\langle g\right\rangle _{\text{{Hake}}}$\tabularnewline
\hline 
\hline 
\multirow{3}{*}{Fall 2009 (AD:T)} & all & $73$ & $0.31\pm0.02$ & $0.39\pm0.03$ & $0.08\pm0.04$\tabularnewline
 & direct-admit & $24$ & $0.35\pm0.03$ & $0.45\pm0.04$ & $0.15\pm0.05$\tabularnewline
 & cond.-admit & $49$ & $0.26\pm0.02$ & $0.30\pm0.02$ & $0.05\pm0.03$\tabularnewline
\cline{2-6} 
\multirow{3}{*}{Spring 2010 (AD:T)} & all & $64$ & $0.33\pm0.02$ & --- & ---\tabularnewline
 & direct-admit & $24$ & $0.40\pm0.04$ & --- & ---\tabularnewline
 & cond.-admit & $40$ & $0.28\pm0.02$ & --- & ---\tabularnewline
\cline{2-6} 
\multirow{3}{*}{Fall 2010 (AD:T)} & all & $56$ & $0.30\pm0.02$ & $0.36\pm0.03$ & $0.09\pm0.04$\tabularnewline
 & direct-admit & $16$ & $0.40\pm0.04$ & $0.60\pm5$ & $0.33\pm0.06$\tabularnewline
 & cond.-admit & $40$ & $0.27\pm0.01$ & $0.26\pm1$ & $-0.01\pm0.02$\tabularnewline
\cline{2-6} 
\multirow{3}{*}{Fall 2010 (Shj:T)} & all & $28$ & $0.23\pm0.02$ & $0.28\pm0.02$ & $0.06\pm0.03$\tabularnewline
 & direct-admit & $6$ & $0.16\pm0.02$ & $0.24\pm0.03$ & $0.10\pm0.04$\tabularnewline
 & cond.-admit & $22$ & $0.25\pm0.02$ & $0.30\pm0.02$ & $0.07\pm0.03$\tabularnewline
\cline{2-6} 
\multirow{3}{*}{Spring 2011 (AD:CWP)} & all & $57$ & $0.26\pm0.01$ & $0.35\pm0.02$ & $0.12\pm0.03$\tabularnewline
 & direct-admit & $0$ & --- & --- & ---\tabularnewline
 & cond.-admit & $57$ & $0.26\pm0.01$ & $0.35\pm0.02$ & $0.12\pm0.03$\tabularnewline
\cline{2-6} 
\multirow{3}{*}{Spring 2011 (Shj:CWP)} & all & $21$ & $0.36\pm0.05$ & $0.45\pm0.05$ & $0.14\pm0.07$\tabularnewline
 & direct-admit & $0$ & --- & --- & ---\tabularnewline
 & cond.-admit & $21$ & $0.36\pm0.05$ & $0.45\pm0.05$ & $0.14\pm0.07$\tabularnewline
\hline 
\end{tabular}

\label{FCI basic analysis}
\end{table*}

\subsection {Student Interviews Post-Instruction \label{ssec:interviews}}

Following the pilot offering of CWP (Spring 2011), 18 of 78 students involved were interviewed, primarily to gather feedback more specific than that typically available from standard course evaluation surveys and to provide evidence as to the reliability of the FCI instrument with the KU population. Interview data was gathered as follows, in an effort to limit bias in the responses arising from the heterogeneous nature of the student population. First, a dossier was developed for each of the 78 students, including data about language ability, course exam and CWP session average scores, FCI scores (pre-test, post-test, and raw gain) and conceptual diagnosis (top-3 misconceptions evidenced on FCI pre-test and top-3 most-changed misconceptions evidenced from pre-/post- test analysis). Second, a Venn Diagram was contructed along two categories: high vs. low FCI pre-test and high vs. low FCI raw gain. Several students from each category (low-pre-low-gain, low-pre-high-gain, etc.) were then invited for interview. In each category, one male and one female student were invited and one high-level ELL and one low-level ELL were invited. In total, 18 students were interviewed and all of the above categories were represented by at least one student from the set. Third, each student was interviewed for approximately 1 hour, with 30 minutes devoted to questions about course management and working in groups and the second 30 minutes devoted to conceptual mechanics questions. The aim of the first 30 minutes was to allow students to explore their feelings to the CWP session itself (format, timing, deliverables, team interactions, etc.). What follows in this section is a synopsis of these interviews. All interviews were audio recorded and transcribed and all interviewers used a common set of talking points to focus their questions. These were:

\begin{itemize}
 \item What were your feelings toward the workshop at the beginning of the semester? Did that feeling change or do you still feel the same way? If it changed, what do you think was the cause? 
 \item What is the most complex or most difficult aspect of the lab activity? Which activity did you enjoy the most? Which activity did you enjoy the least? 
 \item What was it like working with 3-4 other students? How frustrating is group work for you? If it is very frustrating, explain why? What changes would you recommend to teams and group work?
\end{itemize}

In general, students’ initial attitudes toward the CWP instructional strategy were mostly negative, but their opinion improved over a period of about a month as the result of experience and better pre-session preparation. Approximately 75\% used words like ``bad'', ``negative'', ``afraid'', or ``scared'' to describe their feelings and 88\% believed the tasks outlined in workshop session agendas were ``very hard'', ``very difficult'', or ``very stressful''. The most often (80\%) reported cause for negative initial attitudes was a violated expectation of a `recipe' or `verification' lab where instructors provided students with procedures/formulas for students to do/validate. Some representative excerpts include:

\begin{quote}
\emph{``It was not like school, where they gave us a formula and asked us to check if it is true.''}
\end{quote}

\begin{quote}
\emph{``...because I am really worrying about my grades and how I will appear in the class, about my instructor's opinion. So, when I saw the workshop like that, I felt like, `what am I going to do'? ...my friends also said to me that, `don’t worry about the workshop, it will be managed and the instructors will give you the papers...''}
\end{quote}

\begin{quote}
\emph{``We didn't expect that it was this kind of lab because according to students before in this class, it was just following instructions and then you're done.''}
\end{quote}

After 3-4 weeks of experiencing the CWP sessions and having one rotation of team membership, most (72\%) report that their overall feeling toward the workshop session improved and that many (67\%) felt the improvement was the result of ‘just getting used to it’ and adjusting their pre-session preparations such as; reading the text, solving recommended homework problems, and discussing general strategy with team members. Some representative excerpts include:

\begin{quote}
\emph{``After second and third workshops, I started to know how to prepare myself for the workshop.''}
\end{quote}

\begin{quote}
\emph{``...after a few weeks, maybe 4 weeks, I just get used to it. I used to read the book chapters. I said to myself, `I will be OK and will do what I can do'.''}
\end{quote}

\begin{quote}
\emph{``I started to read the [textbook] chapter. I started to discuss the chapter with my friends who had the workshop at the same time. So, if we got the chance to be in one group, we can help each other.''}
\end{quote}

\begin{figure*} 
\includegraphics[bb=0bp 180bp 792bp 612bp,clip,scale=0.55]{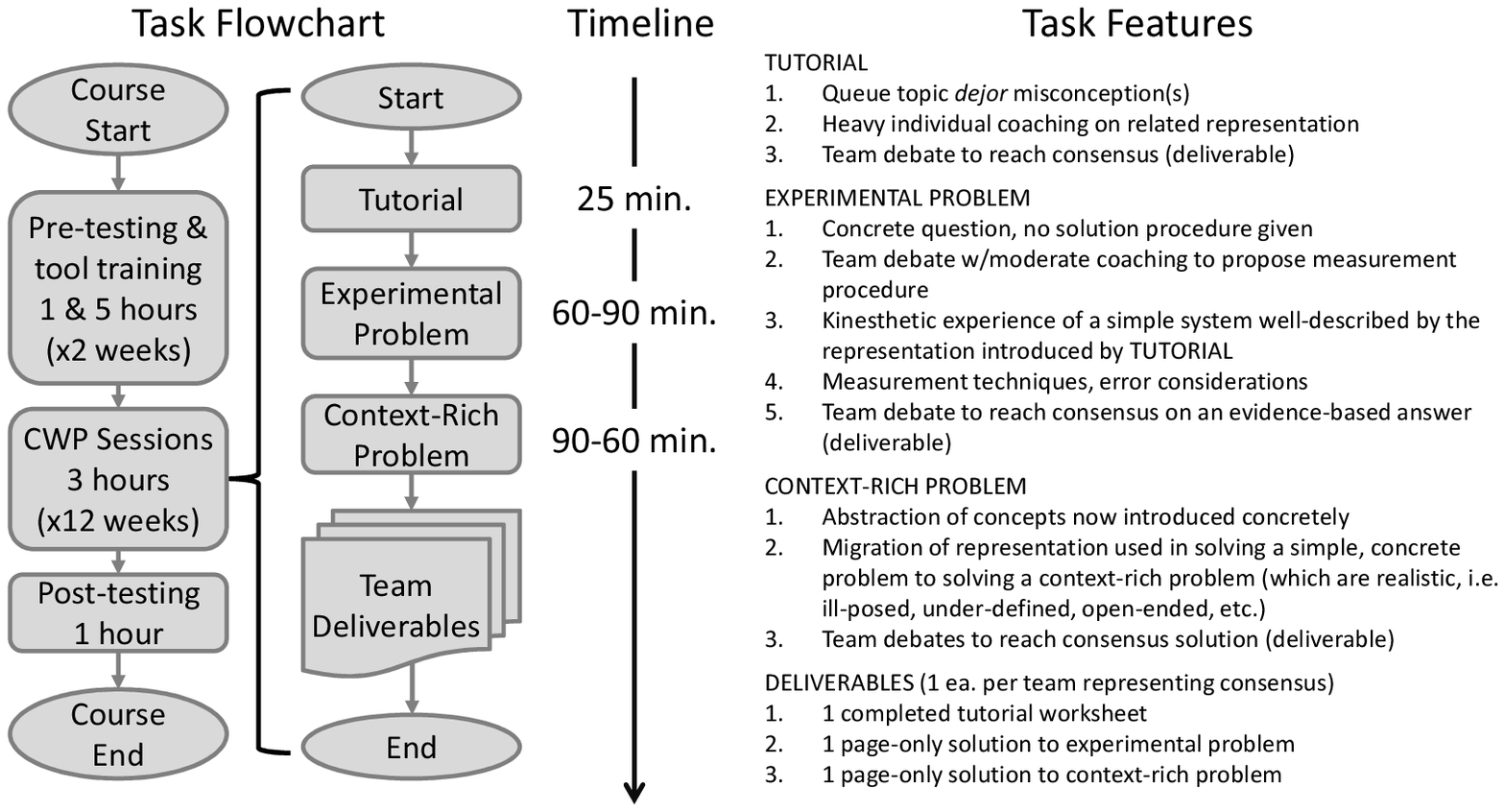}
\caption{A flowchart representation, timeline, and description of tasks in the 3-hour Collaboration Workshop. \label{fig:taskflowchart}}
\end{figure*}

Regarding workshop session elements themselves (see Sec.\ref{sec:designcwp} and Fig.\ref{fig:taskflowchart}), 60\% felt the cooperative group problem was the most difficult and the remaining 40\% felt it was the experimental problem (no one identified the tutorial as the most difficult task). Interestingly, a majority (63\%) of those interviewed also found cooperative groups problems to be the most enjoyable of the three tasks. The remaining 37\% found the experiment problems to be the most enjoyable (no one identified the tutorial as the most enjoyable task). When asked what the least enjoyable task was, 50\% had no opinion. Of the other 50\% of respondents, 38\% said the cooperative group problem was consistently the least enjoyable and 12\% said it was the experimental problem.

Regarding teamwork and peer-interactions, 38\% reported that all group work was frustrating, regardless of group membership or specific details of the learning task. Of these respondents, 100\% said the cause of their frustration was that `other members arrived unprepared'. Another 33\% of these respondents added that `other members didn't do much work' and another 33\% added `other members did not put forward ideas for solutions'. Many respondents included `reading the textbook' as a minimal part of `being prepared' and that advanced reading contributed to their ability to `come up with ideas' during the workshop. Several respondent anecdotes evidenced this situation within teams. For example:

\begin{quote}
\emph{``Like, if [other group members] study hard, then they will be able to give one idea or two ideas in the workshop. Not necessarily all the ideas, but if they don't study and are coming to the lab without studying the chapter already, then you need to be able to do it by yourself, because the others, they didn't study the chapter.''}
\end{quote}

\begin{quote}
\emph{``...and I was asking [other group members], `so, when you read the chapter, what ideas do you have?' And one of them said like, 'Well,... I really didn't read the chapter'.''}
\end{quote}

A separate 37\% of all students interviewed reported that not all group work was frustrating, but that specific teams and/or specific tasks were. Some causes reported by these respondents include, `on a specific activity, other members were unprepared' (65\%), `on a specific team, one person dominated the discussion' (35\%), and `on a specific team, one person did not do their role well' (30\%). The remaining 25\% of respondents either reported no significant frustration (13\%) or declined to comment (12\%). 

When asked to recommend changes to teaming, 50\% of the respondents made specific recommendations, but all of these recommendations had a common desire to remedy perceived inequity in team member roles and responsibilities. Some example comments include:

\begin{quote}
\emph{ ``...I don't like working on a group of 4 other people or 3 other people ... because I know what is going to happen. The whole group will depend on someone, one student.''}
\end{quote}

\begin{quote}
\emph{``...[group work] was really hard. My colleague who was high in physics, he want you to finish it as fast as you could, even if you didn't understand it. Meanwhile, our understanding is, like,... small understanding. It was very hard for us to work with him.''}
\end{quote}

\begin{quote}
\emph{``...give grades depending on the role. For example, manager would get no grade if the group didn't do well, and then the recording and then the skeptic, like, something like that.''}
\end{quote}

\begin{quote}
\emph{``Maybe we can put the good students being all in one group and those that don't work in another group? (long pause) But then maybe half of the workshop would be with groups that may need help, and other groups getting all the marks.''}
\end{quote}

Of the remaining respondents, 25\% of them made no recommendation and 25\% made no recommendation and added a positive affirmation of the teaming recipe used, calling it ``ideal'' or ``necessary'' for the difficulty of the learning tasks assigned to them. Interestingly, even those in this last category often commented, similar to comments presented above, on a dilemma created within groups that is caused by a desire to `get help' and `get grades' by having a highly-skilled peer in the team, but that has the negative side-effect of enabling the tendency of less-skilled members to encourage and exploit dominating behaviors from highly-skilled members. For example, one respondent remarked:

\begin{quote}
\emph{``Just to stick with what [the instructor] did with us, because if you're going to meet certain people in the same group, for example the first workshop, we don't know that that person is clever and is very good in physics, and that [a second] person needs a little bit of help and [a third] person is somewhere between. You don't know anything, right? Because it is the first workshop. So, if you're going to randomly make a group, so then you might collect a group of experts in one group, and another group which,... I can't say they are bad, but they need a little bit of help. So, one group will be perfect. One of them will not receive any help and they will have low marks and other than that, they won't understand.''}
\end{quote}

\section{Collaborative Workshop Physics\label{sec:designcwp}}  

\begin{figure} 
\includegraphics[bb=70bp 120bp 510bp 740bp,clip,scale=0.5]{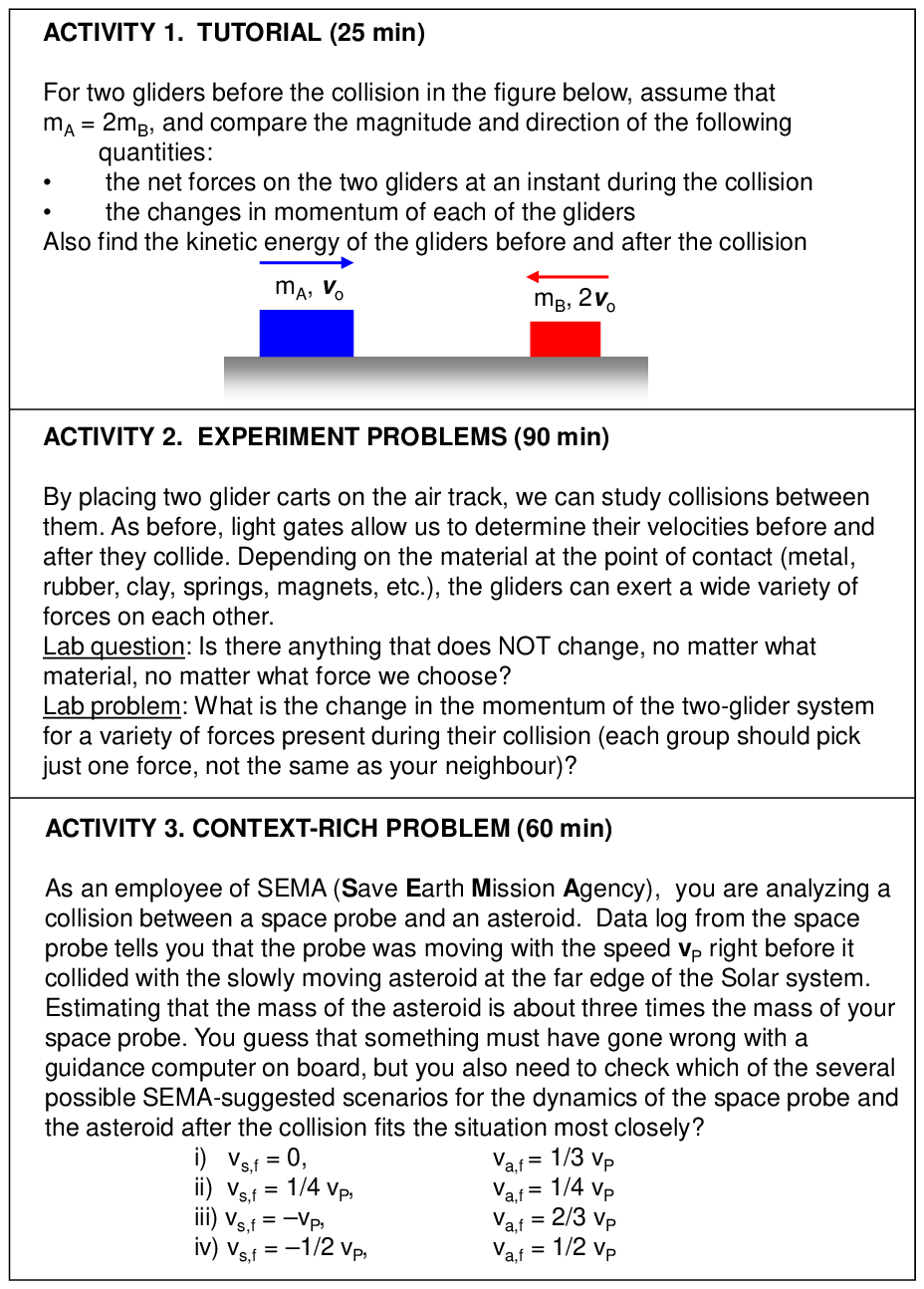}
\caption{An example sequence of activities in a CWP session, in this case, for instruction on linear momentum. \label{fig:exampletasks}}
\end{figure}

In this section, the design and features of our instructional strategy are briefly described. We converged on a hybrid approach which we call ``Collaborative Workshop Physics'' (CWP). A major goal of CWP was to provide `proof-of-principle', to show that IE pedagogy could be successfully adapted to a UAE and Gulf Arab cultural context. The cultural values and expectations of our students were incorporated in the design process and given parity with more typical reform project objectives (e.g. improving conceptual learning gains, providing hands-on activities, etc.) with the goal of minimizing pedagogically unnecessary negative expectancy violation. An exhaustive account of the design and evaluation process followed is available online in a early version of this report\cite{earlyCWP}. In total, 8 instructors were involved in the project, from a variety of cultural (US, MENA, Europe) and institutional backgrounds (US, MENA, Europe). Each session is attended by 25-30 students in teams of 3-4 students each and 2 or 3 instructors serve as facilitator/coaches for the duration of the session. There were no statistically significant differences between their respective classrooms\cite{earlyCWP}.


\subsection{CGPS as a backbone}

Cooperative group problem solving (CGPS) emerged early in our group's considerations, as a likely candidate for best PER-based instructional strategy for the KU context, for a variety of reasons. The design philosophy followed in creating CGPS is stated by Heller \& Heller\cite{CGPS1999} as, ``...a conservative model that conforms with the usual structure and focus of the large introductory physics course...'' and that ``the Minnesota model is based on the familiar triad of lectures, laboratories, and recitation section''. So superficially, a CGPS implementation appears more like a traditional course and avoids confronting students and stakeholders in the reform project with an arguably unnecessary violation of expectations, in terms of the course contact time model. Much like the reform project reported by Goertzen\emph{et al.}\cite{goer2011}, our group avoided reforming the overall course format or the lecture portion in particular, and instead focused our efforts on reforming a weekly 3-hour contact time reserved for the course's traditional verification lab. Furthermore, CGPS requires no reduction in topical content coverage, requires no special rooms, includes its own laboratory curriculum (that can be implemented separately from discussion sessions), and one of these authors (AFI) has prior training and experience in the method. Positively, CGPS' context-rich problems were seen very favorably because of their similarities with learning tasks in innovative engineering design education. Effective design problems in engineering education literature (e.g., Dym \emph{et al.}\cite{dym2005} and references therein) are often described as being ``realistic,'' ``ill-posed,'' or ``open-ended'' which are terms also used to describe context-rich physics problems. Unlike common solution strategies for traditional end-of-chapter problems, both design problems and context-rich physics problems require similar skills like tolerance for uncertainty, estimation, big-picture thinking, self-questioning for clarification, teamwork, and multiple representation use, and they call upon similar cognitive resources and produce similar cognitive loads. This similarity is attractive for a KU reformed physics course for creating a ``knock-on'' effect since all of our students are engineering majors. Perhaps most important, in terms of cultural expectations at KU, the CGPS method has already been studied with gender equity in teaming recipes\cite{hell1992a} and efficacy with under-represented and at-risk student groups\cite{etki1999} in mind and has evidenced positive improvements in DFW rates and learning gains in both studies.

\subsection{Hybridization with KU mini-design-labs and UW Tutorials}

Despite its attractive features, it was neither possible nor preferable to implement canonical CGPS, given the KU context. Figure \ref{fig:taskflowchart} shows a flowchart representation, timeline, and description of tasks in the final form of the 3-hour Collaboration Workshop, as it was piloted in this study, and shows the place of CWP sessions in the overall course structure. The CGPS recitation session was directly implementable, but the laboratory curriculum relies heavily on video cameras \cite{CGPSLabs2007}, equipment that KU does not have. However, since the recitation and laboratory sessions were originally designed to operate independent of each other\cite{CGPS1999}, we chose to create an instructional strategy that borrowed only the recitation techniques. That meant however, that there would be no provision for a lab curriculum to queue mechanics misconceptions in a concrete, kinesthetic manner. To mitigate this, we took inspiration from the ``box-of-probes'' philosophy of \emph{Workshop Physics}\cite{laws1991,WorkshopPhysics} and created simple, open-ended activities from scratch, using equipment already available. We found the simplest way to convert existing recipe lab activities to more conceptually demanding reformed versions was to narrow the experimental goals (shortening time required to 60-90 min.) and removing their given procedures, requiring students instead to \emph{design} the experiment to answer a single, open-ended question. With students, we called these mini-labs \emph{experiment problems} and in hindsight, they are not unlike. They are given to the student teams in the form of a single question and the main tasks are to reach a consensus with their teams on a measurement protocol, execute with the available measuring tools, and answer the question within one written page, using evidence-based reasoning, their measurement results and error analysis. The particular phenomenon investigated is chosen such that it is a concrete experience of a simple system sharing the same underlying physical principles involved in a context-rich group problem, to which it was paired. By posing the experiment problem before the context-rich problem, students' mechanical misconceptions are afforded an opportunity for concrete queuing. Instructor coaching to teams, during their procedure design and when later solving the context-rich group problem, then takes on the form of short `Socratic dialogues' to illicit reflection on and provide targeted teaching interventions. Ultimately, it's left to peer discussions, enriched by such periodic coaching visits, to discover the correct interpretation or solution, and proceed once a team consensus is reached.

We also hybridized our CWP instructional strategy with University of Washington (UW) Tutorials\cite{UWTutorials} and used context-adapted versions of these as introductions for each new concept. Despite the intent to cognitively prepare students for a context-rich problem with the paired experiment problem, this alone was deemed insufficient and particularly risky, in terms of causing negative violations, given KU students' strong expectation for teachers to provide procedures. Therefore, certain UW Tutorials were selected that could serve as a heavily scaffolded training activity with constructing representations, with the representation featured in the tutorial being the one most useful for thinking through the experiment and the context-rich problems. Instructor coaching is more individual and the deliverable (completed tutorial page) is ungraded, so that students have a no-risk opportunity to engage the session's main concept. Together with the three learning tasks; tutorial, experiment problem, and context-rich problem, our intent is to form a routine, prescriptive sequence of tasks, so that as a situation the CWP session feels highly structured and students need only tolerate uncertainty in the individual learning tasks, in small, controlled bursts.

\subsection{Contextually-motivated deviations from canonical cooperative groups}

In addition to hybridization with other PER-based innovations, we also chose to deviate from standard CGPS\cite{CGPS1999}, by changing some aspects of forming and managing students' cooperative groups, for contextual reasons. The differing roles assigned to each member are essentially standard, but one of the most important confounding influences among predictors of learning is student language level with the language of instruction. Conceptual pre-test scores are important, but less so. Therefore, teams were rebuilt at the standard frequency (every 3-4 weeks), and designed such that there is an intentional \emph{heterogeneous skill distribution}, with each team containing at least one member having a relatively high FCI pre-test score, one having high in-class quiz scores (in later weeks), and one having a high IELTS score, to serve as a translator when necessary. More importantly for the KU context, teams are \emph{gender-homogeneous} and so do not match the ideal composition, as determined by Heller \& Hollabaugh\cite{hell1992b} with US students. While in a US context, gender-balanced or even female-majority teams have shown to enrich student-student interaction by mitigating discussion-dominating male student behaviors\cite{hell1992b}, in the KU context, any gender-mixing with KU freshmen students has a paralyzing effect on discussion. We attempted to form such groups, in isolated experiments with volunteers, but after greeting one another, these teams quickly `froze'. There is simply no experience interacting with the opposite gender in a classroom setting and the compounded anxiety of reformed instruction and gender integration is too much to bear all at once. Furthermore, based on FCI pre-test and normalized gain scores for the pre-reform course (see Tab.\ref{demographics}) there is significantly less evidence of a gender-gap compared to US students, and so we felt gender-homogeneous teaming was less of a threat to equity in learning at KU. Furthermore, typical, full-scale CGPS implementations distribute students differently over lab and recitation sections, so that the teams created for the respective situations are drawn from different rosters, but in our case, the same team negotiates all three tasks; tutorial, lab, and recitation, together and in one sitting. Finally, for more institutional reasons, we were unable to ensure individual accountability within groups in the manner recommended by by Heller \& Hollabaugh\cite{hell1992b}. There, the authors compared two different strategies for creating positive interdependence within groups (\emph{goal interdependence} and \emph{reward interdependence}), with the objective of fostering mutual concern for individuals' success within groups and personal accountability to contribute toward the group effort. They found that \emph{reward interdependence}, created by adding a group problem solved in recitation to the score of individual in-class exams, was superior in this regard. We were unable to implement the preferred method because course exams at KU are the domain of the lecture session instructors and we were unwilling to impinge upon that tradition in a first-reform project.

\subsection{Example Sequence of CWP Session Activities\label{sub:CWP example activity}}

Figure \ref{fig:exampletasks} illustrates an example set of activities used in the CWP sessions and helps to explain how they are chosen to form a coherent sequence. In this example, students are working through a variety of tasks revolving around linear momentum and conditions for its conservation. First, notice that in all the tasks, there is little or no numerical information given. This is done to reinforce explicit attention to reading in the problem solving strategy. Students are instructed to keep their pens and pencils down for the first 5 minutes of each activity and to read only. During this time, the instructors make their first round of coaching visits, asking students to reflect on their understanding of the task; ``What is the big idea?'' and ``What does the writer of this problem want from you?,'' ``What is the goal your team needs to reach?'' Feedback and coaching is given on text analysis. For example, many students struggle with the multiple uses of the word ``moment'' and its derivatives (i.e. ``momentum''). Some conclude, quite reasonably, that the word is used in reference to `time' (i.e. a very short time interval), rather than to torque or momentum. This first round of coaching allows instructors to engage in short discussions about context and how context in the problem statement can modify the meaning of jargonized words such as these. If necessary, the class will be stopped for a few words from the instructor if an issue appears to be common to all groups. One instructor recalls in their journal a 5 minute discussion of the word `hammer', for the case when it is used as a verb (i.e. `to hammer a nail into wood') rather than as a noun to identify the tool. The pattern of coaching visits, for all three session tasks, was typically as follows:

\begin{itemize}
\item On the first visit

\begin{itemize}
\item Strong encouragement to \emph{read} problem statement 
\item Discuss, ``What is being asked of us?''
\item Generate a large number of \emph{ideas} for possible solutions
\item Withhold criticism of each other's ideas
\item Strong discouragement to touch any lab equipment or make measurements
for first 20-30 minutes of the problem
\end{itemize}
\item On a typical second visit (10 minutes after discussions begin)

\begin{itemize}
\item Socratic questioning of the team, to gauge and to guide clear understanding
of the problem statement
\item Strong encouragement to begin eliminating the weakest ideas for solutions
\item Strong encouragement to \emph{balance} time spent converging on solutions
versus building an apparatus and conducting measurements
\end{itemize}
\end{itemize}

On third and later visits, instructors focused on questioning the team about features of their chosen solution strategy (e.g. ``Why did you choose this detector over another?) and what would happen if they made modifications (e.g. ``If we change the location of this photogate, what will happen to your graphs?'') Early in the course, during the first 1-2 CWP sessions, there was also often a need to stop the whole class and present a few tips for effective team work. At the conclusion of the experiment problem, teams were often encouraged to take a 5-10 minute break. Upon return, one context-rich group problem is distributed to each Recorder. Again, teams are coached to not write anything for the first 10 minutes or so, but rather to read, discuss, and answer among themselves, ``What is the goal, what is the writer of this problem asking us for?'' Instructors made a visit after this initial period to illicit reflection.

\section{Discussion\label{sec:discussion}}

\subsection{How far away from the context of the developing institution can a PER-based instructional strategy be implemented?\label{ssec:discusshowfar}}
This work suggests, as shown in Fig. \ref{fig:examfcivslang}, that PER-based innovations can be implemented across great differences in cultural context (\emph{d}). However, the care required in selecting or designing an instructional strategy, that will implement well in the adopting context, from PER-based approaches available in the literature is not trivial, as shown by the design process followed to converge on a specific reformed instructional strategy. Our hypothesis was that implementation of \emph{any} PER-based approach would be very difficult due to the culturally differing expectations that the student population brings to the classroom context, relative to the US culture for which PER-based approached were originally optimized and that the difficulty of implementation would be evidenced by relatively lower improvements to conceptual learning gains, exam performance and other measures of learning. In this case study however, problem-solving ability on traditional problems is consistently improved for all student groups and moderately high language proficiency alone explains the difference between improvement to FCI learning gains relative to other SIs, as shown in Figs. \ref{fig:gainvscontext} and \ref{fig:examfcivslang}. In other words, in the present case there does not appear to be evidence of a residual `cultural effect' limiting improvements to learning gains.

\subsection{Can beneficial criteria and failure risks for the reform project be derived from contextual differences between that of the primary and secondary implementing institutions?\label{ssec:discusscriteria}}
In this case study, we find that the context of our course reform project prefigures at least three important beneficial criteria for the reform, in addition to the usual core of PER-based classroom norms. The first is a \emph{prescribed sequence of activities}, as shown in Fig.\ref{fig:taskflowchart}, which is attractive for adoption in the CWP instructional strategy due to KU students' relatively higher aversion for uncertainty (Tab. \ref{tab:nationalculture}), specifically the strong \emph{expectation for highly structured learning situations}, a difference with US student populations which extends to task-level expectations, as shown by pre-instruction MPEX responses on the independence cluster (Fig.\ref{fig:MPEX-pretest}). This issue is an example of the importance of distinguishing task-level and situation-level expectations, for deciding whether or not to accommodate or violate an expectation. A critical PER-based norm is the importance of \emph{sense-making over answer-making} which is in direct conflict with our students' \emph{emphasis on learning `how to do'} over \emph{learning `how to learn'} (see Tab. \ref{tab:nationalculture}), but these both pertain primarily to student reasoning in learning tasks and not necessarily to the larger situation that the tasks are embedded in. For establishing this PER-based norm, it is important to violate expectations to the contrary at the task-level (e.g. using conceptual/qualitative problems, ill-posed problems, experiment design tasks, etc.) but there is no clear reason not to accommodate student expectations at the situation-level in this case and nest the PER-based tasks into highly structure situations with extensive scaffolding. In the post-analysis, student interview data evidences the efficacy of this feature of the CWP instructional strategy, since when asked `What were your feelings toward the workshop?', students unanimously identified individual tasks as the primary cause of anxiety.

The second criteria for our instructional strategy design is use of group-based tasks, but in \emph{gender-homogeneous} teams which is attractive for adoption because of KU students' relatively low individualism, specifically the strong \emph{expectation for grouping based on prior affiliation} (Tab. \ref{tab:nationalculture}) resulting from same-gender grouping throughout secondary-school experiences. Gender-heterogeneous teaming is an example of a situation-level feature, which is common to many group-based instructional strategies in their primary implementations\cite{hell1992b,lorenzo2006,scaleup2007}, but one where the reasons for the feature are not present in the context of our SI. In primary implementations, gender-heterogeneous teams have been a means of fostering rich \emph{student-student interactions}, a critical PER-based classroom norm, in a way that is both diverse and equitable. This in turn has been motivated by a need to address the disproportionate representation, conceptual pre-test and gain performance, and retention of male students in US STEM classrooms. But in the KU context, the representation and retention of female students is comparable to male students (Tab. \ref{demographics}) and the gender-gap in FCI scores is significantly smaller (Tab. \ref{demographics}), perhaps even non-existent. Thus, it appears less likely that gender-heterogeneous teams would be an effective means of diversifying the set of pre-conceptions brought to bear on a learning task and thereby enriching student-student interactions in our context. In fact, in the KU context and especially with freshmen students, gender-heterogeneous teaming certainly has a chilling effect on communication within the group and therefore has the opposite effect on student-student interactions compared to US SI contexts.

The third criteria motivated by our specific context is use of English language level-heterogeneous teams which is attractive for adoption partially because of KU students' relatively high acceptance/preference for large power-distances, specifically the expectation for teacher-lead and teacher-initiated communication (Tab. \ref{tab:nationalculture}). The practical impact of these expectations is to confound with and obscure language-related barriers to comprehension and small-group communication in group-based learning tasks which is a widespread issue at KU since the student population is essentially 100\% ELL (Tab. \ref{demographics}). A student in a team could be quiet and not contributing because she does not understand the task (i.e. low reading and/or low verbal comprehension), because they cannot confidently express their questions or ideas in the language of instruction to their peers, or because they do not expect to be held responsible for initiating communication or any combination of these three. The first two possibilities impacting student expression and learning in physics tasks are supported by correlations seen in traditional problem-solving and FCI data gathered from traditionally taught course offerings prior to the reform project (Fig. \ref{fig:examfcivslang}). From the FCI data, for students scoring below IELTS 6.5 (approx. TOEFL iBT 86), average pre-test and average gain scores are consistent with random response, suggesting that many of these students may be unable to differentiate response options on conceptual multiple choice items. Language-level-heterogeneous teaming was adopted to help accommodate students on the first two possibilities, by providing every team with at least one high-level English language speaker who could help facilitate communication. This consequently better enables instructors to encourage quiet group members to speak out, as it increases the likelihood that any low participation they observe in groups is more the result of cultural expectations and not language-deficiency.

In this case study, we also find the context of our course reform prefigures at least two significant failure risks. The first is the risk posed by \emph{mis-adapting PER-based tasks}. Establishing two of the four critical PER-norms, namely \emph{sense-making over answer-making} and \emph{equity in student-teacher interactions}, is substantially more difficult in the KU context, relative to typical US context. This is anticipated on the basis of student expectations, both at the situational and task-levels of context. The \emph{sense-making over answer-making} norm is arguably the most contrary to expectations, as evidenced by situational expectations related to cultural values of uncertainity avoidance and individualism (Tab. \ref{tab:nationalculture}) where students expect \emph{tasks with sure outcomes that involve following instructions and no risks}, and place an \emph{emphasis on learning `how to do'} over learning `how to learn'. This manifests as a ubiquitous desire of students to know `what is the procedure for the course?', evidenced by indicators (at the task-level) such as the very low pre-instruction MPEX independence cluster scores (Fig. \ref{fig:MPEX-pretest}) and (at the situation-level) by dominantly negative attitudes to open-ended, ill-posed problems in the CWP sessions in post-instruction student interviews (Sec. \ref{ssec:interviews}). This state of affairs places strong and consistent pressure on instructors to ease student anxiety by adapting PER-based tasks (like design labs or cooperative group problems) beyond just matching the local context (i.e. change unfamiliar scenarios to familiar ones, ``police office'' rather than ``state trooper'', replacing references to things like ice-skating, lightning, and rain storms, etc.), to truncating the task so that it can be solved by a rote, formulaic strategy.

The second risk posed to the reform project in the KU context is related to the first and is connected to establishing the critical PER-norm of \emph{equity in student-teacher interactions}. Students in the KU context have a strong expectation, motivated by cultural values (Tab. \ref{tab:nationalculture}) that \emph{teachers should possess and transfer absolute truths}. By its very nature, this belief presupposes that a learning task has one right answer which the skillful teacher knows and the ethical teacher shares. This is reinforced by prior experience, as evidenced in student post-instruction interviews (Sec. \ref{ssec:interviews}), where a large majority of respondents report that they expect lab is `following the procedure' and `validating the physical laws', and that this is how lab has been (in secondary school) and should be (in university) administered. This expectation significantly complicates IE pedagogy because it \emph{de facto} categorizes instructors' efforts to teach reasoning as an effort to `hide the right answer' which a typical student considers unethical and possibly contemplates reporting it as such to university administrators; a serious threat to any reform project.

\subsection{How and to what extent can the original instructional strategy be changed, to both preserve its core functions and accommodate the new context?\label{ssec:discussaccomodate}}
  
As discussed in Sec. \ref{sec:designcwp}, there are some features of the backbone CGPS PER-method that were preserved, despite their potential to cause strong, negative expectancy violations, and some features which were not adopted despite significant associated benefits reported in the PER literature. Again, we find the level of context and its associated expectations important for addressing each choice. The indicators of SI success in the present case suggest that as long as the integrity of PER-based tasks (e.g. context-rich problems, tutorials, etc.) is maintained and there is no compromise on \emph{answer-making over sense-making}, there is a great deal of flexibility allowable in the design of situations in which they are featured. At KU, a single instance of a student team negotiates all three PER-based tasks in a highly structured sequence in the CWP session together and in a single sitting, quite different from the distributed nature of lecture, lab, and recitation activities in a typical, full-scale CGPS implementation. In terms of learning gains, there is little evidence of a difference (for students with comparable language ability), but we believe, on the basis of student expectations, that this added structure aids greatly in managing student anxiety toward reformed pedagogy. Another example in the present case is the choice to form gender-homogeneous teams, contrary to the ideal team composition\cite{hell1992b}, but beneficial given the prior experiences and expectations of KU students.

A change made to the ideal CGPS implementation strategy that also is noteworthy is the institutionally motivated decision to use goal interdependence (requiring team consensus on solutions to learning tasks), rather than reward interdependence (including a cooperative group problem on individual exam scores), to foster mutual concern for success and individual accountability within teams. In their second seminal paper on CGPS\cite{hell1992b}, Heller \& Hollabaugh report, when using the first strategy, that ``In many groups, the students did not take the assignment seriously. They talked about their social life, and rarely finished solving the problem. In other groups, students worked independently to solve the problem, usually using a formulaic strategy instead of the prescribed strategy, then compared solutions''. In our experience, group dysfunctions of this kind almost never occurred. In journals and anecdotes from session instructors, there were no examples of students not taking the problems seriously. As indicated by student interviews, this is likely the result of high anxiety and negative initial attitudes toward the CWP session. Students dreaded the session too much to be dismissive of it. And of instances where students independently solved problems then compared, there were only a few (2-3) reported instances and it was always as a last-resort means of conflict avoidance, when tensions within the group were running high. Instead, student interview data reveals a frequent tension between student attitudes toward \emph{skill-heterogeneous teaming} and \emph{individual accountability} through explicit role assignment and rotation. Half of those interviewed suggested that skill-heterogeneous teaming was necessary for equity, so that no team had a monopoly on members who were perceived to be highly skilled, but that it also created significant instances of dominating behaviors and conflict avoidance. Students who perceived themselves to be less skilled wanted highly skilled teammates, but often felt dominated by such members and avoided disagreeing with them on physical principles in the learning tasks. Conversely, students that perceived themselves to be highly skilled frequently felt that other students did not contribute as much to the group's work. Thus, we find that in the KU context, a qualitative difference with student attitudes and behaviors reported in the Heller \& Hollabaugh study\cite{hell1992b}, is that \emph{skill-heterogeneous teaming} likely increased dominating behaviors and increased conflict avoidance related to the actual physics learning tasks. It is likely that the reward interdependence scheme would have partially eased such tensions. Implementation of the less effective goal interdependence, which caused mild, widespread dysfunction at University of Minnesota, lead to equally widespread but arguably more severe dysfunction at KU. In hind sight, this was one change from the canonical CGPS implementation with clear, negative consequences to the success of the SI in our context.

\section{Conclusion\label{sec:conclusion}}

In conclusion, we find that for students with high English proficiency at KU, normalized gain on FCI improves substantially, from $\left\langle g\right\rangle =0.16\pm0.10$ pre-reform to $\left\langle g\right\rangle =0.47\pm0.08$ (standard errors) in the CWP pilot, indicating a successful SI of PER-based instruction. Furthermore, problem-solving skill was also substantially improved and course DFW rates fell from 50\% to 24\%. This result demonstrates that PER-based instruction can be adapted and successfully implemented in cultural contexts quite different from primary developer institutions in the US. Evidence in post-reform student interviews suggests that prior classroom experiences, and not broader cultural expectations about education, are the more significant cause of expectations that are at odds with the classroom norms of well-functioning PER-based instruction. We also find that contextual differences and culture-specific student expectations, quantified during the baseline characterization of the student population prior to the reform, reliably provided key information for the adaptation and implementation of PER-based instruction. This case study should be valuable for future reforms at KU, the broader Gulf region, and other institutions facing similar challenges involving SI of PER-based instruction outside the US.

This work also raises questions for future research. First, the differential improvements, of students' traditional problem-solving ability (independent of English language ability) versus normalized conceptual gain (only consistent with IE for high English language ability), raises questions about the validity and reliability of FCI with our students, despite making both Arabic and English versions available during testing. Related to this, could lower improvements to normalized learning gain in other SIs also be related to language, or other factors, aside from student expectations? In our perspective, we rely heavily on US culture as a referent because much data exists on US classrooms as the result of PER but, it might be highly informative to conduct a similar study of SIs in nations such as Denmark, Sweden, Great Britain, Ireland, and New Zealand, where the expectations associated with educational situations are more compatible with PER-norms than that of US national culture.


\begin{acknowledgments}
The authors thank the Editor and two anonymous referees with PRST-PER, whose comments and suggestions have greatly improved the manuscript. The authors thank M. Rezeq and J. M. Hassan, for their help gathering FCI and MPEX data in their classrooms. The authors also thank T. R. Khan, R. A. Milne, H. Barada, M. Al-Mualla, A. Al Hammadi, and T. Laursen for their thoughtful review of the original project proposal, and M. Hassan for her efforts to organize an institutional ethics review of the project.

This research is internally supported by the Office of Research Support, Khalifa University.
\end{acknowledgments}

\bibliography{physicseducation}

\begin{thebibliography}{52}
\bibitem [{\citenamefont {Hofstede}(2001{\natexlab{c}})}]{hofs2001uai}%
  \BibitemOpen
  \bibfield  {author} {\bibinfo {author} {\bibfnamefont {G.}~\bibnamefont
  {Hofstede}},\ }\href@noop {} {} (\bibinfo {year} {2001}{\natexlab{c}}),\
  \bibinfo {note} {``Uncertainty avoidance'' is the construct described by G.
  Hofstede as ``society's tolerance for ambiguity''. The expectations listed in
  the text are correlated with these scores and are taken from Figure 4.6, pp.
  169 of G. Hofstede (2001).}\BibitemShut {Stop}%
\end{thebibliography}

\end{document}